\documentclass[USenglish,amsmath]{article}	

\usepackage[utf8]{inputenc}				
\usepackage[big,online]{dgruyter}	
\usepackage{lmodern} 
\usepackage{microtype}
\usepackage[numbers,square,sort&compress]{natbib}
\usepackage{revsymb}
\usepackage{color}
\usepackage{nicefrac}
\usepackage{natbib}
\theoremstyle{dgthm}

\theoremstyle{dgdef}

\newcommand{\ket}[1]{\left| #1\right>}
\def\beq{\begin{equation}}
\def\eeq{\end{equation}}

\begin{document}

	\articletype{Research Article}
	\received{Month	DD, YYYY}
	\revised{Month	DD, YYYY}
  \accepted{Month	DD, YYYY}
  \journalname{De~Gruyter~Journal}
  \journalyear{YYYY}
  \journalvolume{XX}
  \journalissue{X}
  \startpage{1}
  \aop
  \DOI{10.1515/sample-YYYY-XXXX}

\title{Spiraling light: from donut modes to a Magnus effect analogy}
\runningtitle{Spiraling light}

\author*[1]{Robert J.C. Spreeuw}
\runningauthor{R.~Spreeuw}
\affil[1]{\protect\raggedright 
Van der Waals-Zeeman Institute, Institute of Physics, University of Amsterdam,
PO Box 94485, 1090 GL Amsterdam, The Netherlands;
QuSoft, Science Park 123, 1098 XG Amsterdam, The Netherlands; e-mail: r.j.c.spreeuw@uva.nl}
	
	
\abstract{The insight that optical vortex beams carry orbital angular momentum (OAM), which emerged in Leiden about 30 years ago, has since led to an ever expanding range of applications and follow-up studies. 
This paper starts with a short personal account of how these concepts arose. This is followed by a description of some recent  ideas where the coupling of transverse orbital and spin angular momentum (SAM)  in tightly focused laser beams produces interesting new effects. The deflection of a focused light beam by an atom in the focus is reminiscent of the Magnus effect known from aerodynamics. Momentum conservation dictates an accompanying  light force on the atom, transverse to the optical axis. As a consequence, an atom held in an optical tweezer will be trapped at a small distance of up to $\lambda/2\pi$ away from the optical axis, which depends on the spin state of the atom and the magnetic field direction. This opens up new avenues to control the state of motion of atoms in optical tweezers as well as potential applications in quantum gates and interferometry. }


\maketitle

\section{Introduction}

The notion that Laguerre-Gaussian (LG) optical modes carry orbital angular momentum (OAM) of light emerged some thirty years ago \cite{allen_orbital_1992}. This insight came as a surprise even though it was well known 
that light fields must carry angular momentum (AM) determined by their spatial phase distribution \cite{jackson_classical_1999}, in addition to the  better known spin angular momentum (SAM), associated with their polarization. The concept of optical vortices had also been described before \cite{nye_dislocations_1974,coullet_optical_1989}. 
The beauty of LG modes, as well as similar types of vortex beams, is that they provide a particulary clean manifestation of OAM with an integer multiple $l\hbar$ of OAM per photon. The integer number $l$, the  topological charge of the vortex, can be positive or negative, and  arbitrarily large.

These conceptual ideas have since sparked a tremendous amount of activity, branching out to many subfields in physics, both fundamental and applied.
A non-exhaustive sample of follow-up studies  includes the effect of LG modes on the motion of atoms
\cite{babiker_light-induced_1994,allen_atom_1996,lai_radiation_1997}; transfer of OAM to ultracold atoms \cite{tabosa_optical_1999}, Bose-Einstein condensates \cite{andersen_quantized_2006}, and to a bound electron \cite{schmiegelow_transfer_2016}; rotating particles in optical tweezers \cite{he_direct_1995}; creating optical spanner beams \cite{grier_revolution_2003,simpson_mechanical_1997}; 
the use of LG modes to increase the data capacity of optical communication channels \cite{barreiro_beating_2008}; the study of spin-orbit coupling of light in tightly focused beams \cite{bliokh_angular_2010,bliokh_spin--orbital_2011,monteiro_angular_2009,rodriguez-herrera_optical_2010,nieminen_angular_2008}; the generation of  vortex beams of electrons, neutrons, and soft X-rays \cite{verbeeck_production_2010,clark_controlling_2015,lee_laguerregauss_2019}; studying entangled states of OAM beams \cite{mair_entanglement_2001}; generation of ultrafast pulses carrying a controlled self-torque via a high-harmonic generation technique \cite{rego_generation_2019}.

It is not my intention here to give an overview of applications or developments. Several reviews have appeared in recent years, see for example \cite{shen_optical_2019,barnett_optical_2017,pachava_generation_2019,forbes_structured_nodate,franke-arnold_light_2017,franke-arnold_optical_2017,padgett_orbital_2017}.
In this paper I will give a brief personal account of how the concept of OAM first arose in Woerdman's quantum optics group in Leiden. 
This is followed by a discussion of some new ideas with possible applications \cite{spreeuw_off-axis_2020}. 
These  ideas comprise a new  optical analogy of the Magnus effect that pushes a spinning ball on a curved trajectory through the air \cite{magnus_ueber_1853}. 

It should be noted that other optical analogies of the  Magnus effect  have been reported before. These earlier works concerned the rotation of the spatial profile of an optical beam, by coupling to the circular polarization \cite{zeldovich_rotation_1990,dooghin_optical_1992,bliokh_topological_2004,bliokh_geometrical_2006,luo_role_2010,gorodetski_plasmonic_2010}. This effect has been described in terms of Berry phases and is closely related to the spin-Hall effect of light \cite{onoda_hall_2004,hosten_observation_2008,bliokh_geometrodynamics_2008}. 

The analogy discussed here \cite{spreeuw_off-axis_2020} connects to the original Magnus 
effect as viewed in the comoving frame of the rotating ball. In this frame a stream of air particles flows by and is deflected by the spinning ball. Here we replace the ball by a spinning optical dipole in an atom, induced by a focused laser beam. The same beam then gets deflected by this spinning dipole. The focused laser beam thus takes the place of the air stream in the original Magnus effect. 
By momentum conservation the atom will be pushed sideways, just like the rotating ball. This has important consequences for optical tweezers: atoms can be trapped off-axis at a spin-dependent distance from the focus \cite{wang_high-fidelity_2020}.

\section{Birth of an idea}

The first insights about OAM in LG modes---or 'donut modes' as we used to call them---emerged
in the context of studying analogies between classical light and the wave mechanics of a quantum particle. 
Such analogies constituted one broad theme in Han Woerdman's quantum optics research group in Leiden. 
This mode of thinking had been my daily diet during the four years of my PhD work,  exploring analogies between classical optics and two-level  atoms. In late summer of 1991, having just completed my thesis,  I had a few months of time on my hands before leaving for my first postdoc position. Still in the mindset of thinking about analogies, I was entertained and intrigued by the similarities between  Hermite-Gauss (HG) laser mode profiles and  the eigenstates of a 2D  quantum harmonic oscillator (QHO). This is a consequence of a formal equivalence between the paraxial approximation of the Helmholtz equation, and the time-dependent Schr\"odinger equation in (2+1) dimensions, after identifying the propagation direction with the time coordinate. 

In the presence of a quadratic radial refractive index profile (or a sequence of lenses, or convex cavity mirrors), the HG modes would be bound to the optical axis, just like a particle  confined to  a harmonic potential minimum. The optical mode profiles would be identical to the wavefunction of the trapped particle, 
$E_{mn}^\text{HG}(x,y)\propto\psi_{v_x,v_y}(x,y)=\ket{v_x,v_y}$, with the HG mode indices playing the role of the vibrational quantum numbers of the particle in the harmonic trap.

Just like we can form superpositions of QHO eigenstates, we can form the corresponding superpositions of optical modes. In this context, a superposition like $\ket{0,1}+i\ket{1,0}$ is of particular interest because it describes a  particle orbiting around the QHO origin with an angular momentum $\hbar$. 
This observation then raises the question if  the corresponding superposition of HG modes, which would constitute a LG 'donut' mode, could similarly carry angular momentum. 

In support of this thought, the LG modes are invariant under rotation around the optical axis. A rotation is just equivalent to a phase shift, i.e.\ a displacement along the propagation direction. This 
is obvious from the phase factor $\exp[i(kz+l\phi)]$, giving the  wavefront its helical shape. 
The LG modes are eigenfunctions of the rotation operator $\exp(i\phi \hat{L}_z)$ where $\hat{L}_z=-i\partial/\partial\phi$ is the $z$ component of (orbital) angular momentum. The eigenvalues are discrete because the phase  can only change by an integer multiple of 2$\pi$ when going around the optical axis in a closed loop; in the azimuthal phase dependence $\exp(i l\phi)$,  $l$ must be an integer.  
Since this angular momentum is a property of the spatial phase distribution, it is clearly different from the angular momentum as carried by circular polarization. Instead, this is angular momentum of the orbital type, just like electrons can have both orbital ($L$) and spin ($S$) angular momentum.

While in hindsight these notions may seem  obvious, the 
first time I coined the idea of OAM in donut modes, during one of the coffee breaks, it was met with disbelief. It seemed strange that light would somehow move around in orbits. 
Furthermore, conservation of angular momentum  would imply that a donut beam would exert a torque on any absorbing plate, something our intuition was not yet ready to accept.

\section{First checks and early experiments}

Together with Les Allen, who was a guest researcher in the group, we started  some calculations and quickly found that the Poynting vector of a donut beam would spiral around the optical axis. The spiral would be left-handed or right-handed, depending on the sign of the azimuthal mode index $l$. A larger value of $l$ results in a more tightly wound spiral. Thus, if such a beam would fall onto a black disk, there would be an azimuthal component in the radiation pressure on the absorber. The amount of angular momentum was found to be $l\hbar$ per photon. 
Thus the idea started to look more plausible.

As always, Han Woerdman was quick to  ask if and how one could observe the effect experimentally. Could one measure the mechanical torque exerted by a LG beam?
Sending a beam onto an absorbing plate would produce an undesirable amount of heating. A better option seemed to be to use a mode converter made of a pair of cylindrical lenses. Such cylindrical telescopes can modify the phase profile of an optical beam in an astigmatic way, by making use of the Gouy phase. This would allow the conversion of  $l\hbar$ photons into $-l\hbar$ photons without absorbing them. Thus, sending a $l\hbar$ photon through such a convertor would transfer a $2l\hbar$ amount of angular momentum to the cylindrical telescope. For a  laser beam with  power $P$, laser frequency $\omega$, the torque would be  equal to $2lP/\omega$. 

Astigmatic mode conversion also provided a simple technique to convert a HG laser beam into a LG beam. 
The same technique had recently been used independently by Tamm and Weiss \cite{tamm_bistability_1990}. 
In fact, astigmatic mode convertors can be viewed as the OAM-equivalent of quarter- and half-wave plates for SAM (polarization).
With that in mind, an experiment was designed to suspend a cylindrical telescope from a torsional pendulum in  vacuum. 
The idea was essentially to repeat the experiment by Beth \cite{beth_mechanical_1936} which measured the mechanical torque by light due to polarization (SAM). Instead of the quartz waveplate used by Beth, now an astigmatic mode convertor was used. 
In the experiments, conducted by Marco Beijersbergen, the SAM torque as measured by Beth was successfully reproduced. However, measuring the mechanical OAM torque in the same way turned out to be much more prone to strong systematic effects, and prohibitively more difficult. The mechanical torque exerted by a microwave guided mode was in fact successfully measured, although in  this case the torque was a combined effect of SAM and OAM and the two could not be separated \cite{kristensen_angular_1994}.

Other manifestations of the mechanical effects of OAM carrying beams were  observed elsewhere. Absorptive particles were made to spin in the dark center of a TEM$_{01}^\ast$ beam \cite{he_direct_1995}, and an 'optical spanner' was demonstrated \cite{simpson_mechanical_1997}. 
From the earliest conception of OAM onward,  Allen and coworkers maintained a strong interest in the mechanical effects of OAM on atoms  \cite{babiker_light-induced_1994,allen_atom_1996,lai_radiation_1997}. 
Optical OAM was later  transferred to ultracold atoms and Bose-Einstein condensates \cite{tabosa_optical_1999,andersen_quantized_2006}. As we discuss below, mechanical effects can even occur when the incident beam carries no OAM.

\section{Interplay of spin and orbital angular momentum}

While the first concepts of OAM arose in the context of paraxial beams, it is in the nonparaxial regime that the interplay between SAM and OAM becomes interesting \cite{bliokh_angular_2010,bliokh_spin--orbital_2011,monteiro_angular_2009,rodriguez-herrera_optical_2010,nieminen_angular_2008}. Within the paraxial limit, spin and orbital AM of a light mode are essentially additive, they can have  independent good quantum numbers, the angular momentum being $(l+\sigma)\hbar$ per photon.
For  non-paraxial light fields, SAM  and OAM  can still be independently measured but  $l$ and $\sigma$ are in  general no longer good quantum numbers. The total angular momentum $J=L+S$ does remain a good quantum number
\cite{enk_spin_1994,enk_commutation_1994}.  We now discuss some new ideas that make use of this spin-orbit coupling \cite{spreeuw_off-axis_2020}. 

Two non-paraxial examples will illustrate how SAM and OAM are intrinsically intertwined, (i) the field of a tightly focused laser beam, and (ii) the field emitted by a rotating dipole. 
While the former field pattern shows transverse SAM near the focus, the latter shows transverse OAM in the plane of the dipole. The coupling of these  two can produce interesting new effects, in particular the deflection of a tightly focused laser beam by a circular dipole, and off-axis displacement of atoms in an optical tweezer. The effect is  reminiscent of the Magnus effect that pushes  a spinning ball  along a curved trajectory through air \cite{magnus_ueber_1853}. 
Whereas the motion of atoms in OAM-carrying laser beams has been a topic of interest from the early days on \cite{babiker_light-induced_1994,allen_atom_1996,lai_radiation_1997,tabosa_optical_1999,andersen_quantized_2006}, here we consider the situation where the incident beam carries no OAM. Instead, transverse OAM is generated by the circular dipole induced by the laser  beam.

\begin{figure}[t]
\centering
\includegraphics[width=0.9\columnwidth]{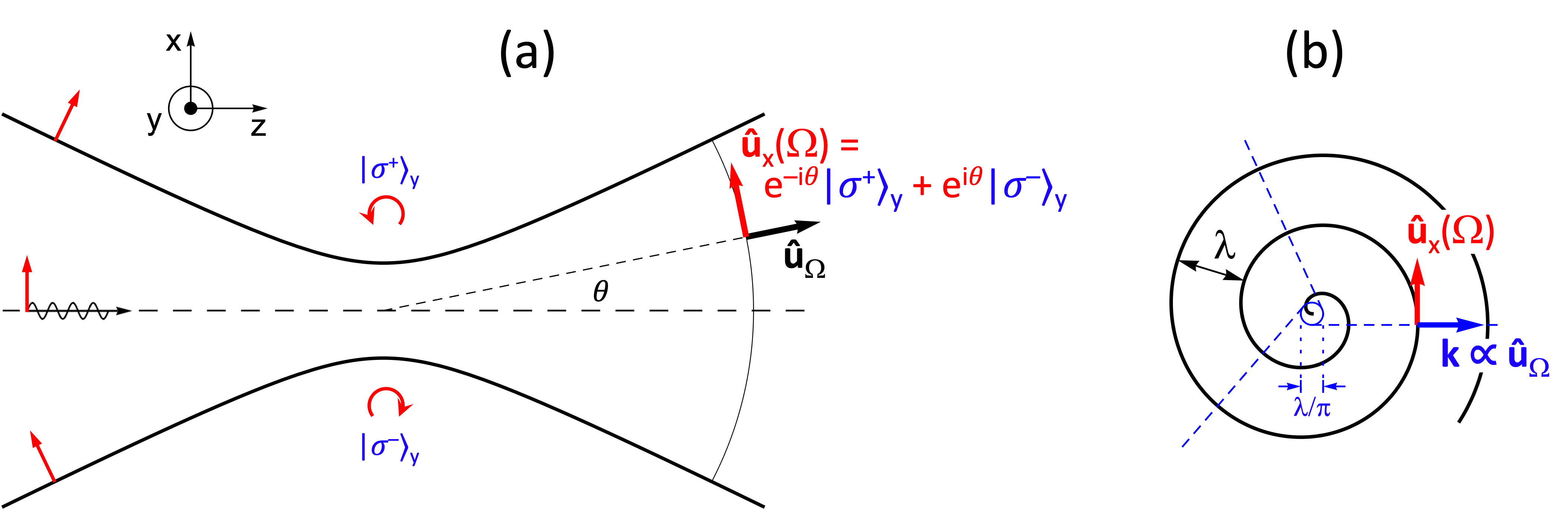}
\caption[]{Examples of non-paraxial light fields with 'intertwined' SAM and OAM. (a) In the wings of a tight focus in a $x$-polarized beam the polarization is circular $(\sigma^\pm)_y$ in the $xz$ plane. Far from the focus the local polarization $\mathbf{\hat{u}}_x(\Omega)\equiv\mathbf{\hat{u}}_x(\theta,\phi)$ is tilted linear which can be decomposed in two spiral waves with opposite handedness ($e^{\mp i\theta}$) and opposite polarization $(\sigma^\pm)_y$ .
(b) In the plane of a rotating dipole, the polarization is in-plane linear, while the  wavefront of the emitted light has a spiral shape. The light appears to originate from a location $\lambdabar=\lambda/2\pi$ away from the atom. For observers in different (in-plane) directions $\mathbf{\hat{u}}_\Omega$, the light appears to originate from a different point $\mathbf{r}=\lambdabar\,\mathbf{\hat{u}}_\Omega\times\mathbf{\hat{y}}$, on a circle with radius $\lambdabar$ (solid). Three example viewing directions are indicated by dashed lines. The apparent origin displacement corresponds to an amount $\mathbf{L}=\mathbf{r}\times\hbar k\,\mathbf{\hat{u}}_\Omega =\hbar\,\mathbf{\hat{y}}$ of transverse OAM per photon, with  $\hbar k\,\mathbf{\hat{u}}_\Omega$ the linear momentum of a photon.}
\label{fig:tifocspiral}
\end{figure}

\subsection{Tight focus}

Let's consider an approximately Gaussian laser beam, $x$ polarized and propagating in the $+z$ direction, with a (tight) waist in $z=0$, see Fig.\ \ref{fig:tifocspiral}(a). In the $xz$ plane the field displays strong field gradients  near the focus,  not only in amplitude but also in polarization \cite{thompson_coherence_2013}. The latter can be seen by recognizing that well before the focus (more than a Rayleigh range, $z\ll -z_\text{R}$)  the wavefronts are spherical surfaces to which the local polarization must be 
parallel. The incident light on either side of the 
optical axis $z$ will then have its  polarization   tilted forward or backward, so that the  local polarization is $\mathbf{\hat{x}}\cos\theta\pm\mathbf{\hat{z}}\sin\theta$. 

In the focal plane, $z=0$, which  is   a flat wavefront,  the tilted polarization components combine to linear $x$ on the optical axis. Away from the axis, however, the  components have different phases. Moving toward the  $+x$ direction the plane-wave component coming from above will be advanced in phase, whereas the component from below will be delayed. The corresponding tilted linear polarization components thus add up to elliptical polarization in the $xz$ plane. 
This means that the field locally carries SAM pointing in the $y$ direction, i.e.\ transverse SAM, which will change sign as one passes the $z$ axis. 

One may now  wonder where this angular momentum came from, considering that the incident beam is simply linearly ($x$) polarized. For this it is illuminating to look again at the polarization far from the focus, for example on a spherical surface large compared to the Rayleigh range, $R\gg z_\text{R}$. A plane-wave component propagating in the direction $\theta$ has a local tilted polarization $\mathbf{\hat{x}}\cos\theta-\mathbf{\hat{z}}\sin\theta$. Decomposition of this linear polarization into its circular components $\mathbf{\hat{u}}_\pm = (\mathbf{\hat{x}}\mp i\,\mathbf{\hat{z}})/\sqrt{2}$ yields 
\begin{equation*}
	\mathbf{\hat{x}}\cos\theta-\mathbf{\hat{z}}\sin\theta=
	\frac{1}{\sqrt{2}}\left( e^{i\theta}\mathbf{\hat{u}}_-+e^{-i\theta}\mathbf{\hat{u}}_+\right)
\end{equation*}
In this expression, the angle-dependent phase factors $e^{\pm i\theta}$ show that the circular field components are arranged on spiral wavefronts, indicating transverse OAM. The combinations are such that  positive SAM ($\mathbf{\hat{u}}_+$; $\sigma_y=1$) is paired with  negative OAM ($e^{-i\theta}$; $l_y=-1$), and vice versa. We have to keep in mind, of course, that these circular components  are not transversely polarized, so they cannot propagate independently of each other.

\subsection{Spiral wave from a circular dipole}

\begin{figure}
\centering
\includegraphics[width=0.5\columnwidth]{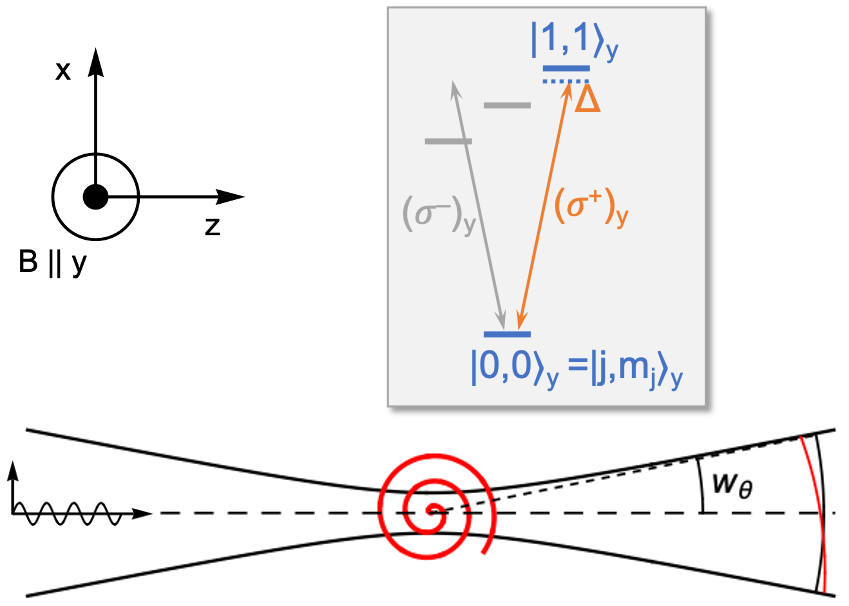}
\caption[]{Optical analog of the Magnus effect. A linearly polarized ($\mathbf{E}\parallel x$), focused laser induces a circular dipole ($xz$ plane) on a $j=0\rightarrow j'=1$ ($\Delta m_j=1$) transition, with a magnetic field $\mathbf{B}\parallel y$ setting the quantization axis. The spiral wave scattered by the circular dipole interferes with the  incident wave. Due to the relative tilt between the wavefronts (exaggerated for clarity), interference may shift the optical power to one side of the beam, thus  deflecting the beam in the $xz$ plane. The corresponding reaction force on the atom, can shift the equilibrium trapping position in an optical tweezer away from the optical axis by an amount $\lambdabar=\lambda/2\pi$, see text and Fig.\ \ref{fig:pdrive}.
}
\label{fig:wftilt}
\end{figure}

As a second example of the interwovenness of SAM and OAM let's consider the field emitted by a rotating dipole (Fig.\ \ref{fig:tifocspiral}(b)), for example in an atom with a $j=0\rightarrow j'=1$ transition. For later use we assume 
a magnetic field  $\mathbf{B}\parallel\mathbf{\hat{y}}$ 
that separates the upper magnetic sublevels $\ket{m_{j'}}_y$ and defines the quantization axis, see Fig.\ \ref{fig:wftilt}. If the $\Delta m_j=1$ transition is driven by laser light with a $(\sigma^+)_y$ polarization component (with respect to the $y$ quantization axis), this can induce a circular dipole, rotating in the $xz$ plane. The light scattered by this rotating dipole will now appear differently to observers from different directions. 

To an observer along the $y$ axis, perpendicular to the plane of rotation, the dipole will simply appear as a rotating dipole, emitting circularly polarized light, i.e.\ carrying SAM. 
An observer in the plane of rotation, on the other hand, will only see the projection of the dipole perpendicular to her viewing direction. The dipole will  appear as an oscillating linear dipole, emitting linearly polarized light in the $xz$ plane (Fig.\ \ref{fig:tifocspiral}(b)). This may seem, naively, to violate the conservation of angular momentum. A  $\Delta m_j=1$ photon must carry away one $\hbar$ unit of AM, so where did it go? The conservation law is restored when we recognize that the in-plane light now carries transverse OAM. 

This becomes clear by noting that a second observer in the $xz$ plane would also observe linear in-plane polarization, with the same amplitude but with a different phase, since the observed projection of the dipole reaches its maximum at a time that depends on the viewing direction.  The phase difference will be  equal to the angle between the two  observation directions, and reveals that the oscillating dipole is in fact rotating. An observer who goes around the dipole in a closed loop will see the  phase increase or decrease by $2\pi$, depending on the sense of rotation of the dipole. Thus, in the plane of the dipole, the wavefront of the emitted light takes the shape of a spiral. 

The field emitted in a direction $\mathbf{\hat{u}}_\Omega=(\cos\phi\sin\theta, \sin\phi\sin\theta, \cos\theta)$ by a $\mathbf{\hat{u}}_\pm$ polarized dipole is proportional to $\left( \mathbf{\hat{u}}_\Omega\times\mathbf{\hat{u}}_\pm\right) \times\mathbf{\hat{u}}_\Omega$ \cite{jackson_classical_1999}. Here we use $\Omega\equiv (\theta,\phi)$ as the pair of spherical angles. 
In the plane of the  dipole ($\phi=0$),
\begin{equation*}
	\left(\mathbf{\hat{u}}_\Omega\times\mathbf{\hat{u}}_\pm\right) \times\mathbf{\hat{u}}_\Omega=\frac{e^{\pm i\theta}}{\sqrt{2}}\left(\cos\theta,0,-\sin\theta\right)
\end{equation*}
the spiral wave character is apparent from the angle-dependent phase factor $e^{\pm i\theta}$. 
It is  this spiral-wave phase factor that gives  the circular dipole pattern transverse OAM in the $xz$ plane.

Compared to a circle, a spiral has  of course a  small local tilt so that the normal to the wavefront does not point to the origin. This peculiarity has already been recognized by C.G.\ Darwin, who stated that for circular dipoles ``\dots the wave front of the emitted radiation faces not exactly away from the origin, but from a point about a wave-length away from it.''\cite{darwin_notes_1932}. With the spiral picture in mind, one quickly sees that this point must lie a distance $\lambdabar=\lambda/2\pi=k^{-1}$ away from the atom. An intriguing detail about this apparent displacement is  that observers from different in-plane viewing angles $\mathbf{\hat{u}}_\Omega$ will disagree about where the source appears to be located. The  apparent  locations, $\mathbf{r}=\lambdabar\,\mathbf{\hat{u}}_\Omega\times\mathbf{\hat{y}}$, form a circle with radius $\lambdabar$ around the dipole, see Fig.\ \ref{fig:tifocspiral}(b). 
A recent observation using a trapped ion showed that a circular dipole is indeed imaged to a location beside itself \cite{araneda_wavelength-scale_2019}. 

Multiplying the displacement by  the momentum of a photon $\hbar\mathbf{k}=\hbar k\,\mathbf{\hat{u}}_\Omega$, we retrieve exactly the amount of $\mathbf{L}=\mathbf{r}\times\hbar k\,\mathbf{\hat{u}}_\Omega =\hbar\,\mathbf{\hat{y}}$ of angular momentum per photon, which is  now of orbital nature.
Thus the angular momentum of light emitted by a circular dipole is entirely spin when viewed on-axis, but entirely orbital when viewed in-plane. In intermediate directions, the total angular momentum would still be $\hbar$ per photon but the OAM and SAM parts would be fractional. 
The dipole pattern as a whole is an eigenfunction of $J_y=S_y+L_y$, but not of $S_y$ nor $L_y$ separately \cite{enk_spin_1994,enk_commutation_1994}. The non-separability of SAM and OAM has been described as a form of spin-orbit coupling in tightly focused laser beams \cite{rodriguez-herrera_optical_2010,bliokh_spin--orbital_2011,bliokh_angular_2010,monteiro_angular_2009}.

\subsection{Circular dipole in a tight focus, spin-orbit coupling}

Let's now combine the two examples above and see what happens when a circular dipole field is scattered in a linearly polarized laser field that excites the dipole. The two effects mentioned above, Magnus-like beam deflection and off-axis tweezer trapping,  are most clearly manifested in slightly different situations, but the calculation is  similar. Therefore, let's first consider the conceptually simplest situation of a  $j=0\rightarrow j'=1$ transition, in the presence of a magnetic field (quantization axis)  $\mathbf{B}\parallel\mathbf{\hat{y}}$, see Fig.\ \ref{fig:wftilt}. 
The field enables the spectral selection of the magnetic sublevels $\ket{m_{j'}}$, by Zeeman shifting them in energy (see below for typical values).
We consider a $x$-polarized  laser beam incident along the $z$ axis. The $\mathbf{\hat{x}}$ polarization can induce a $(\sigma^+)_y$ dipole with its $1/\sqrt{2}$ projection on $\mathbf{\hat{u}}_+$. However it is not correct to think of this process as removing $(\sigma^+)_y$ polarized photons from the incident beam. After all, that would leave the beam with a surplus of $(\sigma^-)_y$ character which cannot propagate in the $z$ direction of the beam. Instead, the atom will scatter a $(\sigma^+)_y$ dipole pattern with AM of partly spin and partly orbital nature.

As shown in Fig.\ \ref{fig:wftilt}, in the $xz$ plane the spiral wave front is slightly tilted with respect to the spherical wave fronts of the forward incident beam. Since this tilt corresponds to a gradient of their relative phase, it may result in constructive interference on  one side of the beam, and at the same time destructive interference  on the other side. This implies  a deflection of optical power toward the constructive side. Since light carries linear momentum, such deflection implies a reaction force on the atom, $F_x<0$ if the beam is deflected towards $+x$.

From the spiral-wave picture we can immediately see that the force will disappear if we displace the atom by an amount $\lambdabar$ to the side of the optical axis, because the tilt between the wavefronts then vanishes, and with it the  beam deflection. In an optical tweezer  the atom will find an equilibrium trapping position at a distance  $\lambdabar$ off-axis. 

Thus, while the emission of a circular dipole {\em appears} to come from  a different position \cite{darwin_notes_1932,araneda_wavelength-scale_2019}, the position of the dipole in an optical tweezer may {\em truly be} different, i.e.\ away from the focus. 
This true displacement of the trapping location can be seen as a counterpart of the apparent displacement of the emitter location \cite{wang_high-fidelity_2020,spreeuw_off-axis_2020}.

\section{Calculation}

The calculation of Magnus-like beam deflection and off-axis tweezer trap displacement was described in Ref.\ \cite{spreeuw_off-axis_2020}; the essentials are summarized here. 
As sketched in  Fig.\ \ref{fig:wftilt}, we place an atom with a $j=0\rightarrow j'=1$ transition in the origin. Using  a magnetic field $\mathbf{B}\parallel\mathbf{\hat{y}}$ we separate the upper magnetic sublevels $\ket{m_{j'}}_y$  by the Zeeman shift $g_{j'}m_{j'}\mu_B B$,  with $g_{j'}$ the Land\'e factor and $\mu_B$ the Bohr magneton. A typical value for the Zeeman shift would be $\sim 100\,\text{MHz}$.
We tune the $x$ polarized incident laser beam near the $\Delta m_j=1$ component, with a detuning $\Delta$ small compared to the Zeeman shift, so that we can neglect the other $\Delta m_j$ components. A typical value  would be $\Delta/2\pi\sim 10\,\text{MHz}$. 
The induced  circular dipole rotates in the $xz$ plane, at the optical frequency of the laser field, $\omega/2\pi\sim 10^{14}-10^{15}\,\text{Hz}$.
For monochromatic light of frequency  $\omega=c k$ the incident field can be written as  $\frac{1}{2}\mathbf{E}_\text{in}(\Omega) e^{-i\omega t}+c.c.$, so that $\mathbf{E}_\text{in}(\Omega)$ is the amplitude of the incident field propagating in the direction $\mathbf{\hat{u}}_\Omega$
with wave vector $\mathbf{k}=k\mathbf{\hat{u}}_\Omega$. Throughout we shall write only the positive frequency components ($\sim e^{-i\omega t}$) of the fields.

For comparison, we consider two different shapes of incident beams, Gaussian (G) and `angular tophat' ($\Pi$), where the latter approximates the output of a uniformly illuminated focusing lens. The field for these two beams can be written as 
\begin{eqnarray}
	\mathbf{E}^\text{(G)}_\text{in}(\Omega) & \approx & \mathcal{E}_0^\text{(G)}\,\exp[-\theta^2/w_\theta^2]\,\mathbf{\hat{u}}_x(\Omega) \label{eq:EinG}\\
	\mathbf{E}^{(\Pi)}_\text{in}(\Omega) & = &
 	\mathcal{E}_0^{(\Pi)}\,\Pi(\theta/2r_\theta)\,\mathbf{\hat{u}}_x(\Omega) \label{eq:EinPi}
\label{eq:EinOmega}
\end{eqnarray}
with amplitudes $\mathcal{E}_0^\text{(G)},\mathcal{E}_0^{(\Pi)}>0$. 
The Gaussian is only an approximate solution of the wave equation because the wings are not strictly zero for $\theta>\pi$. In the paraxial limit its angular width $w_\theta$ relates to the minimum spatial waist  $w_0$ ($1/e^2$ radius of intensity) as $w_\theta w_0=\lambda/\pi$. For the angular tophat, $\Pi(\theta/2r_\theta)$ is the rectangular function with angular half width $r_\theta$ and unit amplitude. Its spatial profile near the focus is the familiar Airy disk pattern. Note  that neither propagation phases nor the Gouy phase are visible here, as the above  expressions are in  angular coordinates. Also note that neither of these incident beams carries OAM along the optical axis.

The polarization unit vector $\mathbf{\hat{u}}_x(\Omega)$ must be transverse to $\mathbf{\hat{u}}_\Omega$; here it is obtained by co-rotating  $\mathbf{\hat{x}}$ when rotating  $\mathbf{\hat{z}}\rightarrow\mathbf{\hat{u}}_\Omega$, i.e.\ rotating by $\theta$ around an axis $\mathbf{\hat{z}}\times\mathbf{\hat{u}}_\Omega$ \cite{richards_electromagnetic_1959,rodriguez-herrera_optical_2010}, 
\begin{equation*}
	\mathbf{\hat{u}}_x(\Omega)=
	\begin{pmatrix}
	\cos\theta  \cos^2 \phi +\sin ^2 \phi  \\
	(\cos \theta -1)\sin \phi  \cos \phi  \\
   	-\sin \theta  \cos \phi 
	\end{pmatrix}
\end{equation*}

The total field is the sum of the incident and scattered waves,
\begin{equation*}
	\mathbf{E}(\Omega)=\mathbf{E}_\text{in}(\Omega)+\mathbf{E}_{\rm{sc}}(\Omega)
\end{equation*}
with $\mathbf{E}_{\rm{sc}}(\Omega)$ the wave radiated by a coherent dipole \cite{jackson_classical_1999}, in angular coordinates, 
\beq
	\mathbf{E}_\text{sc}(\Omega) =
	i\,\mathcal{E}_\text{sc}\,\left(\mathbf{\hat{u}}_\Omega\times\mathbf{\hat{u}}_+\right) \times\mathbf{\hat{u}}_\Omega. 
\label{eq:EscOmega}
\eeq
Taking the dipole radiation to be coherent is essentially a restriction to the low-saturation limit. This is not fundamental, but done here for simplicity. The  dipole is here taken to be circularly polarized   ($\mathbf{\hat{u}}_+$).  
The dipole amplitude is then $\mathbf{p}=p e^{i\chi} \mathbf{\hat{u}}_+$, with  $\chi$ the  phase of the $p_x$ component of the dipole, relative to the local driving field. 
The amplitude of the scattered wave  is $\mathcal{E}_\text{sc}=p\,k^2/4\pi\epsilon_0$, with 
\begin{equation*}
	p=\frac{-\alpha_0\mathcal{E}_0}{i+\Delta/\gamma}.
\end{equation*}
Here $\mathcal{E}_0$ is the amplitude of the incident light in $\mathbf{r}=0$, $\alpha_0>0$ is the polarizability on resonance,  $\gamma$ is the natural half linewidth, and $\Delta=\omega-\omega_0$ is the detuning from the $\Delta m_j=+ 1$ transition.

\subsection{Beam deflection}

For the beam deflection, we   calculate the average propagation direction of the total field
and compare it to the incident field. This can be expressed in the radiant intensity $J(\Omega)=|\mathbf{E}(\Omega)|^2/2Z_0$,
with $Z_0=1/\epsilon_0 c$, so that $J(\Omega)d\Omega$ is the power flowing out of an infinitesimal solid angle $d\Omega=\sin\theta\,d\theta\,d\phi$ around the direction $\mathbf{\hat{u}}_\Omega$.
The total radiant intensity is then the sum of  three terms,
\begin{equation*}
	J(\Omega)  = J_\text{in}(\Omega)+J_\text{sc}(\Omega) +J_\text{if}(\Omega).
\end{equation*}
The interference term
\beq
	J_\text{if}(\Omega)=\frac{1}{2Z_0}\left[\mathbf{E}^\ast_\text{in}(\Omega)\cdot\mathbf{E}_\text{sc}(\Omega)+c.c.\right]
	\label{eq:JifOmega}
\eeq
reflects the assumption of a coherent scattered field, as is the case in the low-saturation limit. In general, if the saturation parameter is finite, the scattered field will contribute an incoherent component to $J_\text{sc}(\Omega)$, which would not appear in $J_\text{if}(\Omega)$.

The deflection of the light beam can be expressed as the change in average wave vector $\delta\langle \mathbf{k}\rangle = 
\langle \mathbf{k}\rangle-\langle \mathbf{k}\rangle_\text{in}$ between the  total (incident plus scattered) and the incident wave, using
\begin{equation*}
	\left\langle \mathbf{k}\right\rangle_\text{in}=k\,\frac{\int \mathbf{\hat{u}}_\Omega\,J_\text{in}(\Omega)\,d\Omega}{\int J_\text{in}(\Omega)\,d\Omega}=k\,\frac{\int \mathbf{\hat{u}}_\Omega\,J_\text{in}(\Omega)\,d\Omega}{P}
\end{equation*}
and similar for  $\langle \mathbf{k}\rangle$, omitting the subscript. The total power $P$ is taken to be equal to the incident power, $P_\text{in}=P$. This assumes (again for simplicity) that  non-radiative decay is absent.

The deflection is entirely determined by the 
interference term $J_\text{if}(\Omega)$. The scattered light itself does not contribute, due to the symmetry of the dipole radiation pattern, $J_\text{sc}(\theta,\phi)=J_\text{sc}(\pi-\theta,\pi+\phi)$, so that $\int \mathbf{\hat{u}}_\Omega J_\text{sc}(\Omega)\,d\Omega=0$. For the deflection we therefore have
\beq
	\delta\langle \mathbf{k}\rangle = 
	\langle \mathbf{k}\rangle-\langle \mathbf{k}\rangle_\text{in} =
	\frac{k}{P}\,\int \mathbf{\hat{u}}_\Omega\,J_\text{if}(\Omega)\,d\Omega
	\label{eq:deltak}
\eeq
and for the force on the atom, by momentum conservation,
\begin{equation*}
	\mathbf{F}=-\frac{P}{\omega}\,\delta\langle \mathbf{k}\rangle=-\frac{1}{c}\,\int \mathbf{\hat{u}}_\Omega\,J_\text{if}(\Omega)\,d\Omega
\end{equation*}
While this expression does include the forward radiation pressure force, in the cases of interest here the main force will be transverse to the optical axis, $\mathbf{F}\approx F_x \mathbf{\hat{x}}$. Then (approximately) $\delta\langle \mathbf{k}\rangle\perp \langle \mathbf{k}\rangle_\text{in}\approx k\mathbf{\hat{z}}$ and the deflection angle is 
\begin{equation*}
	|\delta\theta|\approx
	\frac{|\delta\langle \mathbf{k}\rangle|}{k}
\end{equation*}
We will choose $\delta\theta>0$ if $F_x<0$.

Inserting  Eq.~\eqref{eq:EscOmega} and Eq.~\eqref{eq:EinG} or \eqref{eq:EinPi}  into Eq.~\eqref{eq:JifOmega}, the interference term contains the  amplitude product $\mathcal{E}_0^\text{(G)}\mathcal{E}_\text{sc}$
or $\mathcal{E}_0^{(\Pi)}\mathcal{E}_\text{sc}$. 
In the low-saturation limit, the ratios $\mathcal{E}_\text{sc}/\mathcal{E}_0$ can be conveniently obtained by requiring energy conservation \cite{spreeuw_off-axis_2020}. The interference term $J_\text{if}(\Omega)$ then becomes proportional to the total power, and the deflection angle independent of power.

\begin{figure}[t]
\centering
\includegraphics[height=0.28\columnwidth]{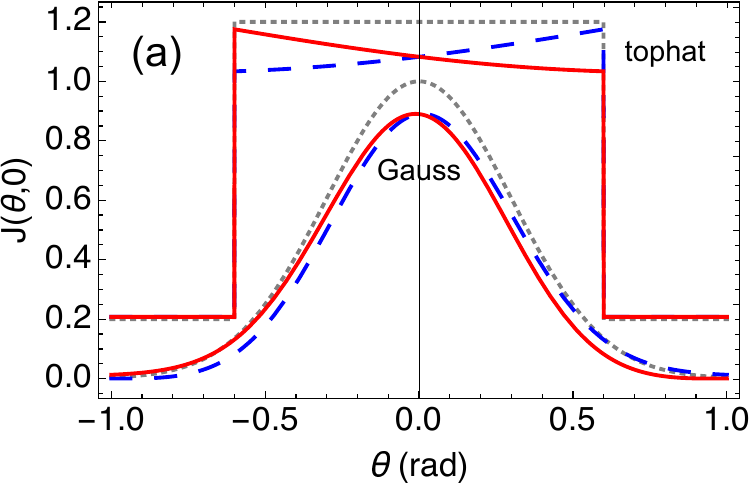}
\hspace{0.05\columnwidth}
\includegraphics[height=0.28\columnwidth]{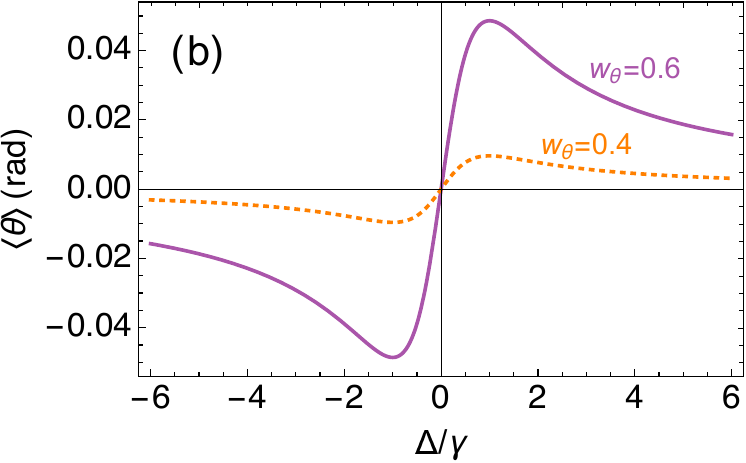}
\caption[]{Beam deflection or Magnus effect analogy. (a) radiant intensities in the plane of the $\mathbf{\hat{u}}_+$ dipole,  for a Gaussian incident beam with $w_\theta=0.6$, and an angular tophat incident beam with $r_\theta=0.6$. The tophat curves have been raised by 0.2 for clarity. 
In both cases, the gray/dotted curve shows $J_\text{in}(\theta,\phi=0)$ of the incident beam, normalized to 1 for $\theta=0$; red/solid and blue/dashed curves show the outgoing, or total  $J(\theta,0)$, for $\Delta=-\gamma$ and $+\gamma$, respectively. For clarity, we identify $(\theta,0)\equiv(-\theta,\pi)$. Curves remain the same upon switching simultaneously the signs of the detuning and the spin of the dipole. (b) the deflection angle of a Gaussian beam as a function of detuning, for two different values of the angular waist $w_\theta$. }
\label{fig:magnuseffect}
\end{figure}

Fig.~\ref{fig:magnuseffect}  shows $J_\text{in}(\Omega)$ in the plane of the dipole ($\phi=0$), together with the total radiant intensity $J(\Omega)$.
For the Gaussian beam, the effect of $J_\text{if}(\Omega)$ is to shift the peak and the average of the direction of propagation away from $\theta=0$. For the angular tophat, the interference leads to an intensity gradient across  the angular  width of the beam, whereas the edges stay at the same angle. In this case the intensity gradient leads to a change in average beam direction.

The corresponding deflection angle is obtained by performing the integration in Eq.~\eqref{eq:deltak},
{\renewcommand{\arraystretch}{1.5}
\beq
	\delta\theta\approx
	\frac{3}{4}
	\frac{ \gamma  \Delta}{\left(\gamma ^2+ \Delta^2\right)}\times
	\left\{\begin{array}{ll}
   		w_\theta^4 & \quad\text{(Gauss)}\\
		r_\theta^4/4 & \quad\text{(angular tophat)}
	\end{array}\right.
	\label{eq:deflangle}
\eeq} 
The results are given as the leading order in $w_\theta$ and $r_\theta$.
The deflection angle  reaches maximal values of $\delta\theta=\pm 3w_\theta^4/8$ and  $\pm 3r_\theta^4/32$, respectively, for $\Delta=\pm\gamma$; it vanishes in the plane-wave limit, $w_\theta, r_\theta\rightarrow 0$.

\subsection{Off-axis trapping in tweezers}

From the deflection angle, Eq.\ \eqref{eq:deflangle},  the reaction force follows as
\beq
	F_x\approx -\frac{P}{c}\delta\theta
	\label{eq:Fx}
\eeq
We recognize in the  detuning dependence that the force is essentially a dipole force \cite{gordon_motion_1980}, arising from polarization gradients  near the focus of a linearly polarized light beam 
\cite{bliokh_spin--orbital_2011,dorn_focus_2003,
monteiro_angular_2009,nieminen_angular_2008,thompson_coherence_2013,wang_high-fidelity_2020,caldwell_sideband_2020}. 
This transverse force will push the atom away from the optical axis. If this happens inside an optical tweezer, the atom will find a new equilibrium trapping position, a small distance away from the optical axis: the tweezer  traps the atom  `where the focus is not'. 

While the size of the displacement is not immediately obvious from Eqs.\ (\ref{eq:deflangle},\ref{eq:Fx}), the argument of tilted wavefronts given above predicts that the atom will be trapped off-axis by an amount $\lambdabar$ in the $x$ direction. Remarkably, the size of the displacement is independent of the detuning, the beam divergence angle, the trap frequency, or even the precise shape of the beam (Gauss vs.\ angular tophat). This profound insight simply follows from the geometric properties of the  scattering problem. 
The simple geometric argument is confirmed  by a calculation (see supplementary material in \cite{spreeuw_off-axis_2020}), that shows that Eq.~\eqref{eq:deflangle} for the beam deflection is multiplied by $1\mp kd$, for a $\mathbf{\hat{u}}_\pm$ dipole displaced by $d$ in the $x$ direction, to lowest order in $d$. Thus the transverse force indeed vanishes for a transverse displacement of $d=k^{-1}=\lambdabar$ in the $x$ direction.

For the transverse forces in an optical tweezer we thus have two equivalent pictures. The first  is a  local one and describes the force in terms of the local intensity gradient of the circular polarization components \cite{thompson_coherence_2013,wang_high-fidelity_2020}. 
The  force can then be calculated in terms of vector and tensor polarizabilities, combined with polarization gradients near the focus. 
The second picture is global/geometrical, according to which the force is determined by interference of the scattered light with the incident light. 
The geometric picture avoids calculation of the local spatial distribution of the light field, as required by the local picture. Instead, it does  need a $k$-space representation of the fields, and performance of an angular integral over the $4\pi$ solid angle. The angular information can be more accessible, especially for non-Gaussian beam profiles.

\section{Experimental considerations}

The deflection of a laser beam ('Magnus effect') and the off-axis displacement of an  atom in an optical tweezer become important in different regimes of physical parameters. For the case discussed so far the atom couples selectively to the $\mathbf{\hat{u}}_+$ component of the light due to the Zeeman splitting in the excited state. This limits the detuning to values small compared to the Zeeman shift, which is not the regime where optical tweezers typically operate. This limitation is easily overcome in different level schemes which will then allow far off-resonant operation. 
We now address the question of what are the best conditions for observing either effect, and discuss a few possible applications.

\subsection{Beam deflection}

From  Eq.~\eqref{eq:deflangle} we see that  the angle of deflection by a single atom is small compared to the divergence angle, $|\delta\theta|\ll r_\theta, w_\theta$.
A direct observation  will thus require sufficiently high signal-to-noise ratio, similar to what was achieved in the recent observation of apparent $\lambdabar$  displacement of an emitter \cite{araneda_wavelength-scale_2019}. 
Furthermore, it  will be necessary to work near resonance ($\Delta\approx\pm\gamma$) to obtain maximal signal. This  however implies that  the photon scattering rate will be relatively high. A  $j=0$ ground state is then a good choice, because it avoids optical pumping between spin states.
On the other hand, near resonance  is not a favourable regime to operate an optical tweezer. A better approach would  therefore be to  hold the atom in an independent trap, such as an ion trap or an additional, tight, far off-resonance optical tweezer. 
To observe the actual beam deflection one could then use a separate,  weak, near-resonant probe beam.

A larger deflection angle may be obtained if multiple atoms cooperate. For example, one may consider dense clouds  of sub-wavelength size, containing tens to hundreds of atoms, that have been observed to show collective scattering properties \cite{pellegrino_observation_2014,machluf_collective_2019}. 
Another possibility may be to use elongated, (quasi-) one-dimensional samples with tight ($\lesssim\lambdabar$) radial confinement, achievable, e.g.,  in optical lattices \cite{moritz_exciting_2003,paredes_tonksgirardeau_2004,kinoshita_observation_2004} and on atom chips \cite{jacqmin_sub-poissonian_2011}. 
Very interesting recent work has shown enhanced optical cross section by the collective scattering of properly spaced arrays of atoms \cite{rui_subradiant_2020,bettles_enhanced_2016,facchinetti_storing_2016,shahmoon_cooperative_2017}.

\begin{figure}[t]
\centering
\includegraphics[width=0.4\columnwidth]{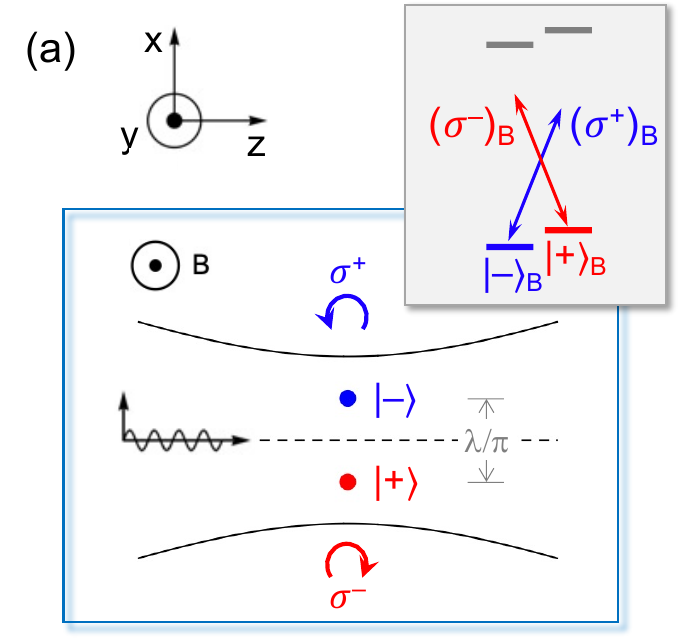}
\hspace{0.1\columnwidth}
\includegraphics[width=0.4\columnwidth]{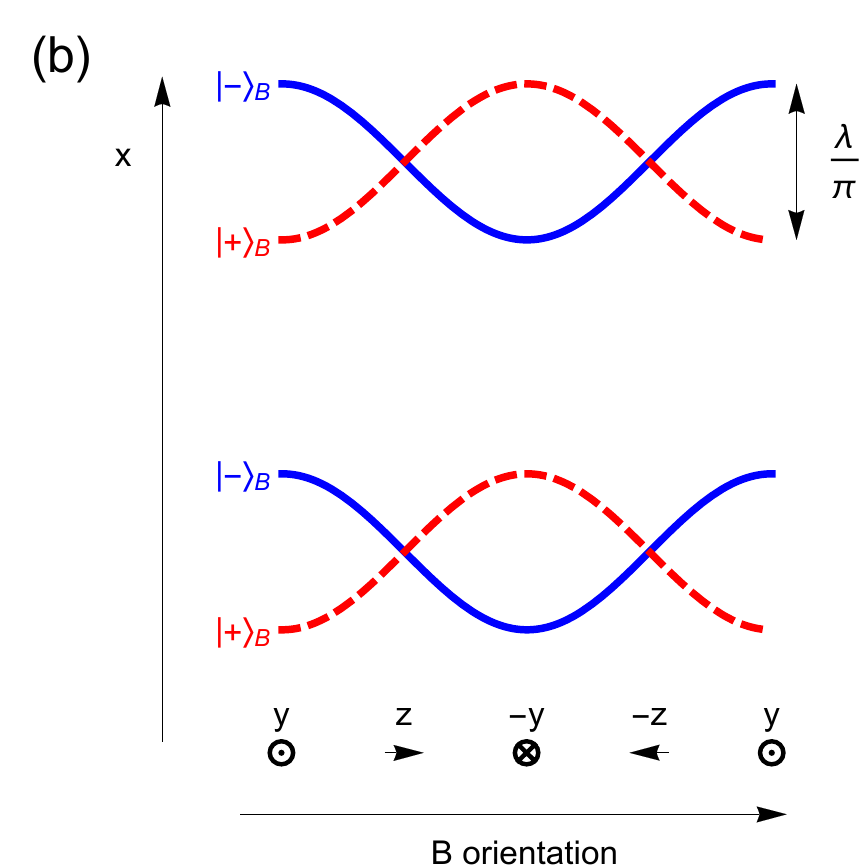}
\caption[]{(a) Optical tweezer operating on a $j=\nicefrac{1}{2}\rightarrow j'=\nicefrac{1}{2}$ transition, leading to $\pm\lambdabar$ off-axis displacements for the $\ket{\mp}_B=\ket{(m_j)_y=\mp\nicefrac{1}{2}}$ sublevels. (b) Spin-dependent motion of atoms in the tweezer, effected by a rotation of the quantization axis ($\mathbf{B}$) through cycles of $y\rightarrow z\rightarrow -y\rightarrow -z\rightarrow y$. The locations of the $\ket{\pm}_B$ traps move up and down along the $x$ axis, in antiphase. The figure shows the situation of  two separate, closely spaced optical tweezers. This may open up possible applications in interferometry, or,  in the presence of  distance-dependent interaction, quantum gates. 
If  $\mathbf{B}$ is rotated at the trap frequency, spin-dependent oscillatory motion in the tweezer can be induced. }
\label{fig:pdrive}
\end{figure}

\subsection{Off-axis trapping in optical tweezers}

For the off-axis displacement of the trapping position in an optical tweezer, the near resonant regime is not suitable, because the  high photon scattering rate produces a large heating rate. 
This can be avoided using a different level scheme, the simplest perhaps being a $j=\nicefrac{1}{2}\rightarrow j'=\nicefrac{1}{2}$ transition. 
Selection rules ensure that the  $\ket{m_j=\pm\nicefrac{1}{2}}_y\equiv \ket{\pm}_B$ states only couple  to the $(\sigma^\mp)_y$ components of the light field
and therefore experience opposite forces $F_x$. The selective coupling no longer requires the energy separation by the Zeeman shift, and remains true even in far off-resonance light, as long as we stay far from other transitions in the atom. 

In this configuration there is no need for a separate probe beam \cite{wang_high-fidelity_2020}, the far off-resonance light ($\Delta/2\pi\sim1-10\,$THz)  of the tweezer itself is sufficient. The  photon scattering and associated heating rates can thus be kept as low as in typical tweezer experiments. 
In this case we do assume that the Zeeman shift is large compared to the trap depth $U_0$ (for example $\mu_B B/h\sim10\,$MHz, and $U_0/h\sim1\,$MHz.)

The above argument based on  the relative tilt of the forward wavefronts again leads to a displacement by $\lambdabar$ in the $x$ direction. The $\ket{-}_B$ sublevel will therefore find an equilibrium position in the tweezer at a displaced off-axis location $x_\text{eq}=\lambdabar$. The $\ket{+}_B$ sublevel will have the opposite displacement, so that for the $j=\nicefrac{1}{2}\rightarrow j'=\nicefrac{1}{2}$ transition:
\begin{equation*}
	x_\text{eq}=-2(m_j)_y \lambdabar
\end{equation*}
The tweezer thus traps the atom off-axis, `where the focus is not', in a spin-dependent location. The two  spin components are trapped with a   Stern-Gerlach type separation \cite{wang_high-fidelity_2020,stellmer_detection_2011}. 

Many available atomic level systems can potentially display off-axis tweezer trapping. For example, in $^{88}$Sr the transition 
${}^3P_2\rightarrow {}^3S_1$  provides a $j=2\rightarrow j'=1$ structure. The outer $(m_j)_y=2\;(-2)$ state couples only to the $\sigma^-\;(\sigma^+)$ polarization component, so its spatial shift will be $-\lambdabar\;(+\lambdabar)$.  Similarly, in $^{87}$Rb one could 
operate a tweezer red detuned to the  $D_1$ line (795~nm), driving the two hyperfine lines  $F=2\rightarrow F'=1,2$, again displacing the outer $(m_F)_y=2\;(-2)$ states by $-\lambdabar\;(+\lambdabar)$, as long as the detuning stays small compared to the fine structure splitting of the $D$ lines. It is worth noting  that the use of the magnetic field avoids optical pumping into  dark states or coherent population trapping.

\subsection{Applications, outlook}
  
The off-axis trapping  locations offer  interesting opportunities to manipulate the motion of atoms in the tweezer, see Fig.~\ref{fig:pdrive}. Let us imagine an atom trapped in the $\ket{+}_B$ state. As we  slowly rotate the magnetic field in the $yz$ plane,  the orientation of the spin will adiabatically follow the rotating quantization axis. After rotating the field $y\rightarrow z\rightarrow -y$, the spin will have maintained its orientation relative to $\mathbf{B}$, i.e.\ $\ket{m_j=\nicefrac{1}{2}}_B\rightarrow\ket{m_j=\nicefrac{1}{2}}_B$. However, its orientation will have flipped in space,  $\ket{m_j=\nicefrac{1}{2}}_y\rightarrow\ket{m_j=-\nicefrac{1}{2}}_y$, since  $\mathbf{B}$ has changed direction. The space-referenced spin flip implies that the atom must have moved to the other side of the optical axis. 
The $\ket{\pm}_B$ counterparts move in opposite directions, of course. 
It has been shown that the spatial separation of the $\ket{\pm}_B$ states lifts the orthogonality 
of their spatial wavefunctions \cite{wang_high-fidelity_2020}, which can enable microwave transitions between motional states in a tweezer trap. This opens up interesting opportunities for interferometry. 
The spin-dependent displacements in a rotating field are shown in Fig.\ \ref{fig:pdrive} for two neighboring tweezers. In combination with a distance-dependent interaction, for example based on Rydberg excitation, this may allow a novel type of quantum gate between two qubits in neighboring tweezers.

It may also be interesting to rotate the field at the resonant frequency $\omega$ of the trap, effectively shaking the traps back and forth: $x_\text{eq}=-2(m_j)_B\lambdabar\cos\omega_B t$. The $m_j=\pm \nicefrac{1}{2}$ levels are shaken with opposite phase. Resonant shaking, $\omega\approx\omega_B$,  will then induce an oscillatory motion in the trap, equivalent to a harmonic driving force $F_x=m\omega^2\lambdabar\cos\omega_B t$. 
For a tweezer with a laser wavelength of $\lambda\approx0.8\,\mu$m, a Gaussian waist of 2$\,\mu$m, holding an atom of mass $m=88u$ in a  20$\,\mu$K deep trap, the trap frequency will be  $\omega\approx2\pi\times7\,$kHz.  In a simple driven harmonic oscillator model the atom would acquire enough energy to leave the trap after  only 3.5 drive cycles, corresponding to a velocity of $\sim 6\,$cm/s. 
This could be used to measure the trap frequency by measuring trap loss. One could also switch off the drive at a moment that the spin states move apart at maximal velocity, and switch off the trap at the same moment. After a time of flight one would then image the spin states in different locations, thus yielding a Stern-Gerlach type measurement of spin composition \cite{stellmer_detection_2011}. 
For the amplitude of the rotating magnetic field a few gauss should be sufficient, to ensure that the Larmor frequency is large compared to the trap frequency. Rotating the field at frequencies of $\sim10\,$kHz is well possible, being comparable to what is used in time-averaged, orbiting
potential traps \cite{petrich_stable_1995}.

While the discussion in this paper has focused on atoms as the spinning dipole, the effects should not be 
restricted to atoms. One may ask, for example, whether one could observe them with nanoparticles, much smaller than the wavelength of light. A key requirement would be the preferential scattering of one circular polarization component over the other. The nanoparticles do not have to physically rotate at the optical frequency, only a rotating  electric dipole must be induced. One might think about using  magnetized particles, lined up in a magnetic field, or dielectric particles with a strong Faraday rotation (Verdet constant), again in a magnetic field.

\section{Discussion}

While analogies can be tremendously helpful in guiding one’s thoughts, there are always limits. 
The optical analogy to the Magnus effect  is no exception in that it will break if  we push it too far. It would be tempting to associate the rotation direction of the  dipole with the sign of the deflection angle. In the conventional Magnus effect this is indeed correct: the air stream is deflected in the ball's spinning direction. In the optical analogy, on the other hand, the beam can be deflected in either direction, depending on the detuning from the atomic resonance, see Eq.\ \eqref{eq:deflangle}. The optical case is an interference effect, which is absent in  a stream of air particles. Thus, while the optical case requires coherence, the conventional Magnus effect occurs in an inherently dissipative setting. In fact, the ball's rotation rate will slow down as a result of air viscosity. By contrast, the spinning dipole discussed here is driven by the laser field; it would not spin without it.

Another striking difference is that the air stream in the conventional Magnus effect is usually taken to be uniform (in the absence of the spinning ball), whereas the optical analog vanishes in the plane-wave limit. In this light it is remarkable that the displacement of the trapping position in an optical tweezer is always $\lambdabar$, independent of the size of the waist. Now, if we increase the tweezer waist $w_0$ to an ever greater value, we do eventually end up with a  plane wave, although the displacement is an ever smaller fraction of the waist, $\lambdabar/w_0\rightarrow 0$. Finally, in the plane wave limit, the tweezer no longer confines the atom, so that `trap displacement' loses its meaning. What does remain is the apparent position shift of the circular dipole, as can be observed by imaging the atom, even using plane waves. 
The universality of the $\lambdabar$ displacement also suggests that it may be interesting to investigate this problem from a topological perspective. 
Earlier work has also connected previous versions of the optical Magnus effect with Berry phases and the Aharonov-Bohm effect \cite{bliokh_topological_2004,bliokh_geometrical_2006,gorodetski_plasmonic_2010}.

As a final remark, it would be interesting to generalize the effects for different level schemes. This would include larger values of $j,j'$, as well as different types of transitions, such as magnetic dipole, electric quadrupole \cite{schmiegelow_transfer_2016}, etc. The latter, for example, supports more tightly wound spiral waves $\sim e^{2i\theta}$, which will presumably double the magnitude of beam deflection and trap displacement.

\section{Summary}

A brief personal, historical account of the  days that saw the emergence of OAM has been presented. After that, some new ideas were discussed related to the coupling of transverse spin and orbital angular momentum. It is predicted  that a circular dipole can deflect a focused laser beam, similar to a spinning ball deflecting a stream of air in the Magnus effect. For an atom trapped in an optical tweezer this may lead to a spin-dependent, off-axis  displacement of up to $\pm\lambdabar$. This displacement is  independent of many trap parameters. An external magnetic field can be used to induce spin-dependent motion or to perform Stern-Gerlach type analysis of the spin states of the atom in the tweezer.

\section{Acknowledgments}
I would like to thank  N.J. van Druten, R. Gerritsma, J. Minar, and A. Urech  for  stimulating  and encouraging discussions. This work was supported by the Netherlands Organization for Scientific Research (NWO).


\end{document}